\documentclass[letterpaper,aps,pre,twocolumn,superscriptaddress,groupedaddress,showpacs]{revtex4}  
\usepackage{graphicx}  
\usepackage{dcolumn}   
\usepackage{bm}        
\usepackage{amssymb}   

\def\tab{\;\;\;\;\;\;\;\;}

\begin{document}




\title{Probability distributions for directed polymers in random media with correlated noise}
\author{
  Sherry Chu and  Mehran Kardar \\
  \textit{Department of Physics, Massachusetts Institute of Technology, Cambridge, Massachusetts 02139, USA}
}       
\date{\today}

\begin{abstract}
The probability distribution for the free energy of directed polymers in random media (DPRM) with uncorrelated noise in 
$d=1+1$ dimensions satisfies the Tracy--Widom distribution.
We inquire if and how this universal distribution is modified in the presence of spatially correlated noise. 
The width of the distribution scales as the DPRM length to an exponent $\beta$, in good (but not full)
agreement with previous renormalization group and numerical results.
The  scaled probability is well described by the Tracy--Widom form for uncorrelated noise, but becomes symmetric
with increasing correlation exponent.
We thus find a class of distributions that continuously interpolates between Tracy--Widom and Gaussian forms.
\end{abstract}

\maketitle


The Tracy--Widom (TW) distribution was originally introduced in connection with the probability for the largest eigenvalue
of a random matrix~\cite{tw96}.
It has since acquired iconic status~\cite{quanta} due to applications ranging from bioinformatic sequence alignments~\cite{mn05} to
aircraft fault detection~\cite{Ch12}.
Like the Gumbel and Gaussian distributions, TW is universal in being independent of various underlying (microscopic) details.
However, whereas it is known how the addition of fat-tailed random variables  modifies a Gaussian to a L\'evy distribution,
corresponding limitations for TW are not known.
We take up this question in the context of directed polymers in random media (DPRM)~\cite{kz87,hhz94,hht15}, one of
the more highly studied systems in the TW class~\cite{ic12,qs15}.

The DPRM problem considers configurations of a directed path (no overhangs) traversing a random energy landscape.
Unlike the traveling salesman problem (which allows overhangs and loops), the optimization problem can be solved
in polynomial time with a transfer matrix formalism~\cite{kz87,hhz94,hht15}.
The optimal energy path (or the free energy at finite temperature) exhibits sample to sample fluctuations, which scale
with the path length $t$, as $t^\beta$.
In 1+1 dimensions, and for uncorrelated random energies, the scaled probability of these 
fluctuations satisfies the TW distribution~\cite{ic12,qs15}.
It is known, however, that the exponent $\beta$ is modified if the random energies have long-range (power-law) 
correlations~\cite{mhkz89}.
We examine energy fluctuations in such correlated energy landscapes, and inquire if and how the TW form changes along
with the exponent $\beta$.

As one of the simplest random processes described by the Kardar-Parisi-Zhang (KPZ) equation~\cite{kpz86,mhkz89}, DPRM has been
extensively studied over the past three decades~\cite{kmb91a,kmb91b,hh91}, with renewed recent 
interest~\cite{hh12-13,hhl14,gldbr} due to its connection to TW. 
It is closely related to the Eden~\cite{e61,m87,rhh91}, the restricted solid-on-solid (RSOS)~\cite{kk89,kkal91}, 
and ballistic deposition (BD) models~\cite{v59,m87}.
(Extensive reviews from both statistical physics~\cite{hhz94,hht15} and mathematical~\cite{qs15} perspectives provide an excellent 
background on the subject.)
In the continuum limit, the partition function $W({\bf x},t)$ of a polymer of length $t$ terminating at a point ${\bf x}\in \mathbb R^d$ satisfies
the stochastic heat equation
\begin{equation}\label{eq:StochasticHeat}
\partial_t W({\bf x},t) = \nu \nabla^2 W({\bf x},t) + \eta({\bf x},t) W({\bf x},t)\,,
\end{equation}
where $\nu$ is related to the polymer line tension, and $\eta({\bf x},t)$ is the random energy at $({\bf x},t)$.
[The Cole-Hopf transformation, $W = \exp[(\lambda/2\nu) h]$, maps the above to the KPZ equation,
$\partial_t h({\bf x},t)= \nu \nabla^2 h+\lambda (\nabla h)^2/2+\eta({\bf x},t)$.]

In $d=1$ dimension, the KPZ equation with uncorrelated energies [$\eta(x,t)$, independent white noise at each $x$]
has exact exponent $\beta=1/3$~\cite{k87,ss10,hh12-13}. 
Recently, the exact limiting end-point energy distribution has also been obtained~\cite{j00,ps00}:
The extremal path from the origin to $(x,t)$ for given $x$ and $t$ (called the pt-pt model), 
related to stochastic growth in a radial geometry, obeys TW Gaussian unitary ensemble (TW-GUE) statistics; 
the extremal path from the origin to any $x$ and a given $t$ (called the pt-line model), 
related to stochastic growth in a flat geometry, obeys TW Gaussian orthogonal ensemble (TW-GOE) statistics~\cite{tw96,hhl14}. 
It is known that the exponent $\beta$ can be modified by introducing noise that is fat-tailed 
[$P(\eta) \sim 1/\eta^{1+\mu}$ as $\eta \rightarrow - \infty$]~\cite{z90,jk91,bbp07}, or long-range correlated~\cite{mhkz89}.
The former was considered in Ref.~\cite{gldbr}, concluding that for $0 < \mu < 5$, both the 
scaling exponent and the end-point distributions are inconsistent with the KPZ/TW universality class described above,
but did not focus on the nature of the modified distributions.
Here, we consider the latter, expanding on earlier work in Ref.~\cite{hhl14}.

In the generalization of the KPZ equation proposed in Ref.~\cite{mhkz89}, the random energies are  spatially correlated such that  
\begin{equation}\label{eq:SpatialCorr}
\langle \eta(x,t) \eta(x',t') \rangle \sim |x - x'|^{2\rho-1} \delta(t-t').
\end{equation}
A one-loop dynamical renormalization group (RG) calculation~\cite{fns77,mhkz89} predicts
\begin{equation}\label{eq:DRGb}
\beta(\rho) = \left\{ \begin{array}{ll} 1/3, & 0 < \rho < 1/4, \\ (1+2\rho)/(5-2\rho), & 1/4 < \rho < 1. \end{array} \right.
\end{equation}
Eq.~(\ref{eq:DRGb}) was also obtained in the field-theoretic works of Ref.~\cite{ftj99} and Ref.~\cite{kcdw14}, using a stochastic Cole-Hopf transformation and a nonperturbative RG approach, respectively.
On dimensional ground, the case of uncorrelated noise [$\delta(x)\sim 1/|x|$] corresponds to $\rho=0$, in the regime
where the RG result coincides with the exact value of $\beta=1/3$. 
The limit $\rho=1$ corresponds to the interface of a two-dimensional Ising model in random fields.
The case of $\rho=1/2$ is of particular interest:
The DPRM problem is trivial if the noise does not depend on $x$, in which case the addition of random variables at different
$t$ would lead to a Gaussian distribution whose width grows with $\beta=1/2$ (as predicted by the above).
However, as we shall elaborate below, the numerical procedure used generates non-trivial correlations for $\rho=1/2$ which vary
logarithmically with $|x-x'|$.


We simulate the discrete pt-line DPRM on a square lattice, with random energies on each site $\eta(x,t)$.
The path is directed along the diagonal, such that the minimal energy is
calculated recursively according to the (transfer matrix) relations
\begin{equation}\label{eq:TM}
E(x,t) = \min \{ E(x,t-1) + \eta(x,t), E(x-1,t-1) + \eta(x,t) \}.
\end{equation}
The square lattice is wrapped around a cylinder of size $L$, corresponding to periodic boundary conditions 
along the $x$-direction.
For $\rho < 1$,  random energies correlated as in Eq.~(\ref{eq:SpatialCorr}) are generated using the 
Fourier transform method proposed in Ref.~\cite{mhss96}. 
(A similar method for generating correlated noise was developed in Ref.~\cite{pyhh95}.)
For $\rho = 1$, the noise is constructed as a Brownian bridge, shifted to have zero mean. 
The simulated system size is $L=10^6$, evolved over $t=10^4$ time steps, 
and averaged over $10^2$ realizations.

With finite $L$, the variance of the minimal energy is expected to satisfy the scaling form
\begin{equation}\label{eq:scaling}
\Delta E = \langle (E - \langle E \rangle)^2 \rangle^{1/2} \sim L^\chi f\left(t\over L^z\right) 
\sim \left\{ \begin{array}{ll} t^\beta, & t \ll L^z, \\ L^\chi, & t \gg L^z \end{array} \right. \,, 
\end{equation}
where angular brackets indicate averaging over different realizations (or independent segments in
the same realization) of random energies. We have introduced the scaling function $f$, whose argument
depends on the {\it dynamic exponent} $z$, which quantifies the ratio of scaling exponents in the $x$ and
$t$ directions. The validity of the scaling form requires that the {\it roughening exponent} $\chi$ 
satisfy the identity $\chi=\beta z$. 
In fact, according to dynamical RG, the Galilean invariance of the KPZ equation implies an
additional exponent identity  $\chi + z = 2$~\cite{fns77,mhkz89}, such that there is only one
independent exponent [e.g. $\beta(\rho)$].

We extract the growth exponent $\beta$ and the dynamic exponent $z$ from the collapse of the curves 
of $\Delta E/t^{\beta}$ vs. $L/t^\zeta$ for different times $t$, where $\zeta = 1/z$.
Alternatively, $\beta$ and $z$ can be deduced respectively from the scaling of the energy 
fluctuations $\Delta E \sim t^\beta$, and the transverse fluctuations 
$\Delta x = \langle (x - x_0)^2 \rangle^{1/2} \sim t^{\zeta}$, where $x_0$ is the origin of the polymer. 
This method yields exponents which are in good agreement with the data collapse approach 
for small $\rho$, where finite size effects are less important.

\begin{figure}
\includegraphics[scale=0.385]{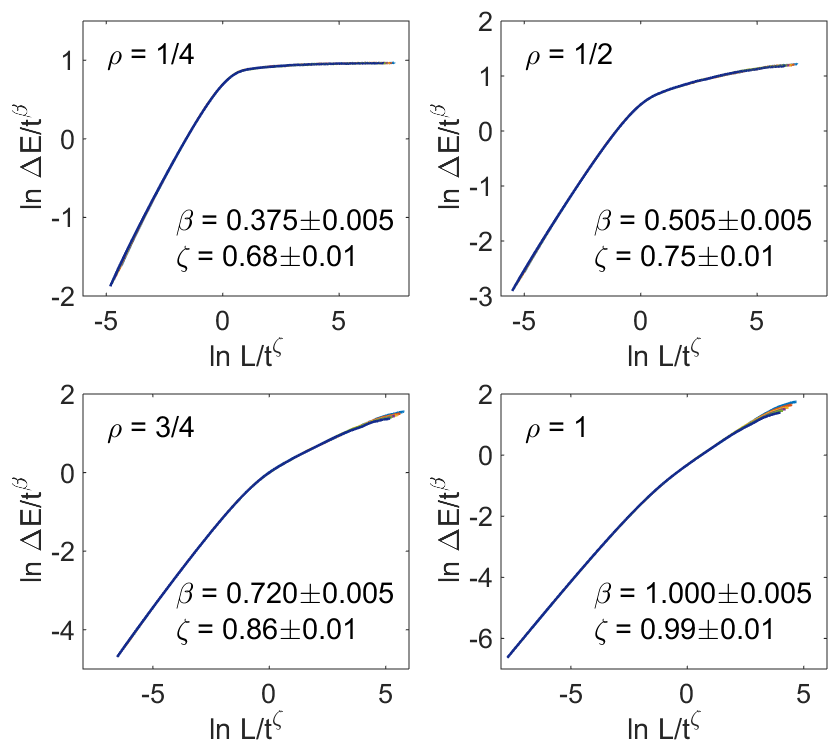}
\caption{Collapse of energy fluctuations of DPRM with a spatially correlated landscape. 
The data corresponds to system size $L = 10^6$, evolved to time $t = 10^4$.
The error bars on the exponents reflect statistical errors in the fits; neglecting potentially larger systematic errors.}
\label{fig: collapse}
\end{figure}

\begin{figure}
\includegraphics[scale=0.385]{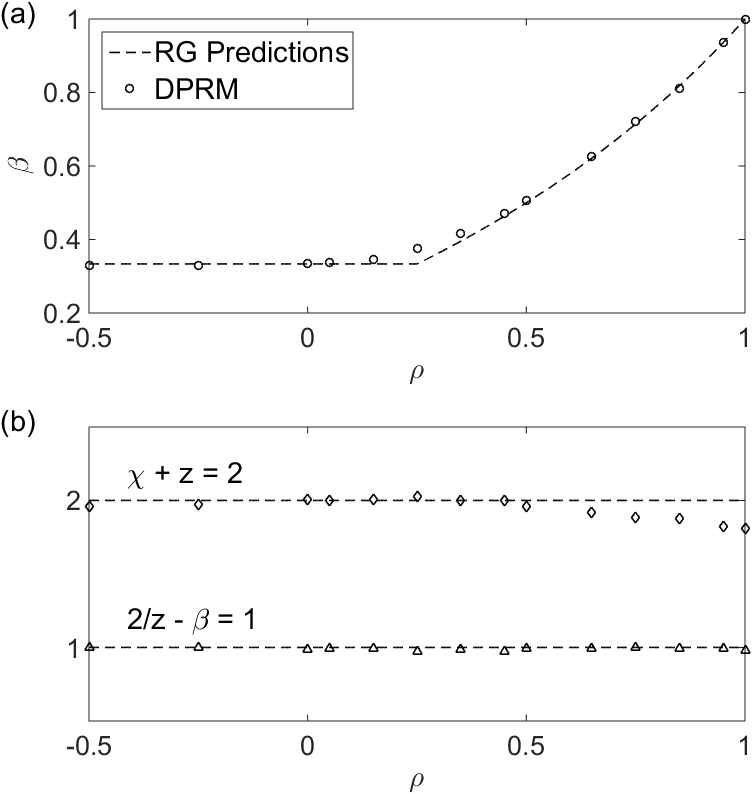}
\caption{{(a)} Variation of $\beta$ with the exponent $\rho$ of spatially correlated energies.
There is a small, but clear deviation from the predicted RG exponents (dashed line).
{(b)} Validity of the exponent identities predicted by Galilean invariance; the discrepancies
are likely a measure of systematic errors.}
\label{fig: exponents}
\end{figure}

As presented in Fig.~\ref{fig: collapse}, the data is very well collapsed according to Eq.~(\ref{eq:scaling}),
although somewhat less so for larger values of $\rho$.
In particular we note the excellent collapse at $\rho=1/4$ which according to the RG result of Eq.~(\ref{eq:DRGb})
is the limiting point for which $\beta$ sticks to 1/3. However, we find $\beta = 0.375\pm0.005$,
(and $\zeta = 0.68\pm0.01$) 
in contradiction to RG, but consistent
with previous results in Ref.~\cite{pyhh95} of $\beta = 0.364\pm0.005$ and $\zeta = 0.692\pm0.005$.
Indeed, as depicted in Fig.~\ref{fig: exponents}, the exponent $\beta$ appears to vary continuously
with $\rho$, in contradiction to Eq.~(\ref{eq:DRGb}).
As in Ref.~\cite{pyhh95}, we extend the simulations to $\rho\leq 0$, and throughout this regime 
obtain $\beta = 1/3$ consistent with uncorrelated noise.
(We also find $\zeta = 2/3$ in this regime  through data collapse as in Fig.~\ref{fig: collapse}). 
For larger values of $\rho$, the agreement with RG improves, and the expected random field Ising
exponents of $\beta = \zeta = 1$ are recovered for $\rho=1$.
The continuous variation of $\beta$ for $\rho \leq 1/2$ is similar to observations in previous 
simulations of DRPM, RSOS, and BD models~\cite{alf91,phss91,wb95,pyhh95}.
We note that the RG exponents are constrained to be exact for uncorrelated noise due to a 
fluctuation-dissipation condition. 
The exponents in Eq.~(\ref{eq:DRGb}), however, follow from a non-renormalization of correlated
noise amplitude, which in view of the numerics is perhaps questionable.

In principle, the scaling relation, Eq.~(\ref{eq:scaling}), involves two exponents ($\beta$ and $\zeta$, or $\chi$ and $z$).
We estimated the roughening exponent $\chi$ from the slope of the collapsed curve in the regime $t \gg L^z$.
A hallmark of the KPZ equation (even with correlated noise) is Galilean invariance~\cite{fns77,kpz86},
which implies the exponent identity $\chi+z=2$. 
The explicit check of this identity presented in Fig.~\ref{fig: exponents} appears to indicate its breakdown
for $\rho>1/2$. However, simply dividing this identity by $z$, and noting $\beta=\chi/z$, yields
a second form $2/z - \beta = 1$, which is excellently obeyed by the data!
The discrepancy between these two identities is an indication of the systematic errors afflicting the fits,
such as the small but systematic curvature in the initial rise of the collapsed curves in Fig.~\ref{fig: collapse},
whose slope is used to obtain the exponent $\chi$.

\begin{figure}
\includegraphics[scale=0.385]{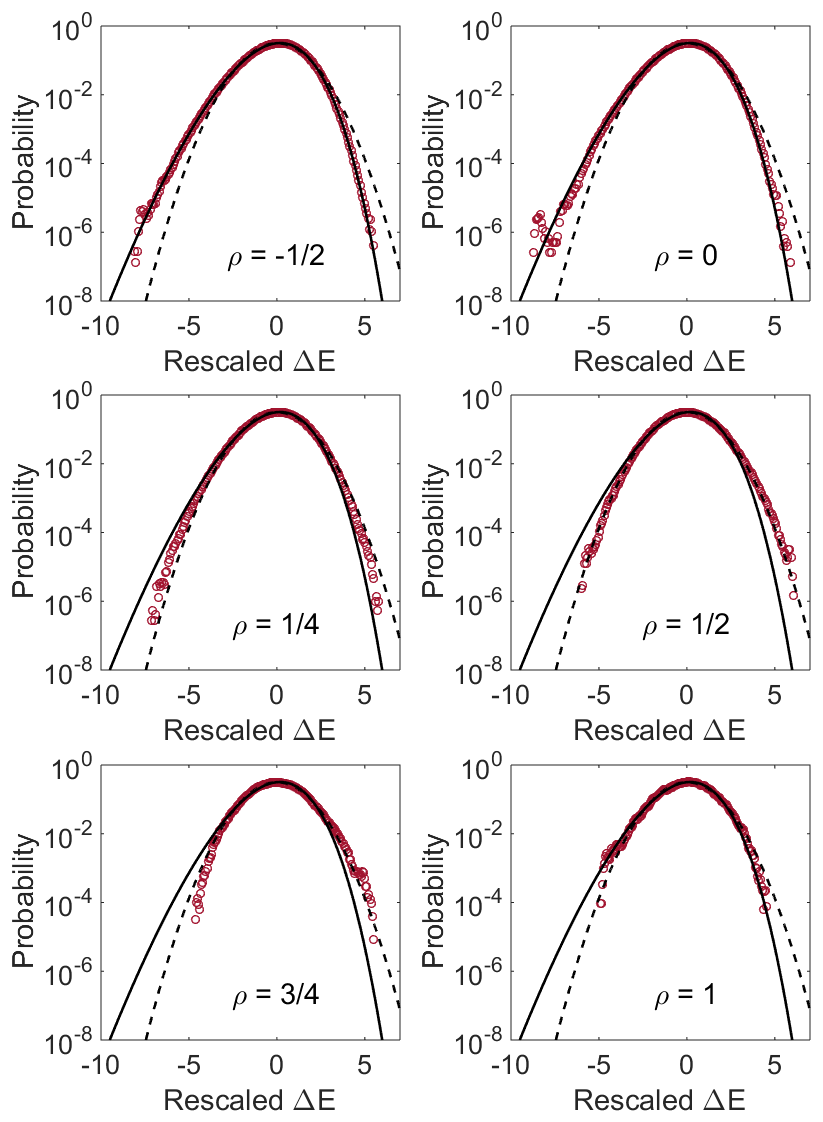}
\caption{Probability distributions for the optimal energy of DPRM for different correlation exponents $\rho$.
The data corresponds to system size $L = 10^6$ at time $t = 5 \times 10^3$, 
rescaled to have mean 0 and variance 1. Results are consistent with the TW-GOE form (solid line)  
for $\rho \leq 0$, and shift smoothly towards Gaussian (dashed line) at $\rho = 1/2$.}
\label{fig: pdfs}
\end{figure}

The end-point energy probability distributions are obtained from time $t=10^3$ to $t=10^4$, 
in increments of $\Delta t = 10^3$, and are normalized to have mean 0 and variance 1. 
The full distributions presented in Fig.~\ref{fig: pdfs} are qualitatively similar to the TW-GOE form
for $\rho \leq 0$, but shift smoothly towards Gaussian as $\rho$ increases to $\rho = 1/2$. 
Beyond $\rho = 1/2$, it is unclear whether the distribution remains Gaussian. 

\begin{figure}
\includegraphics[scale=0.385]{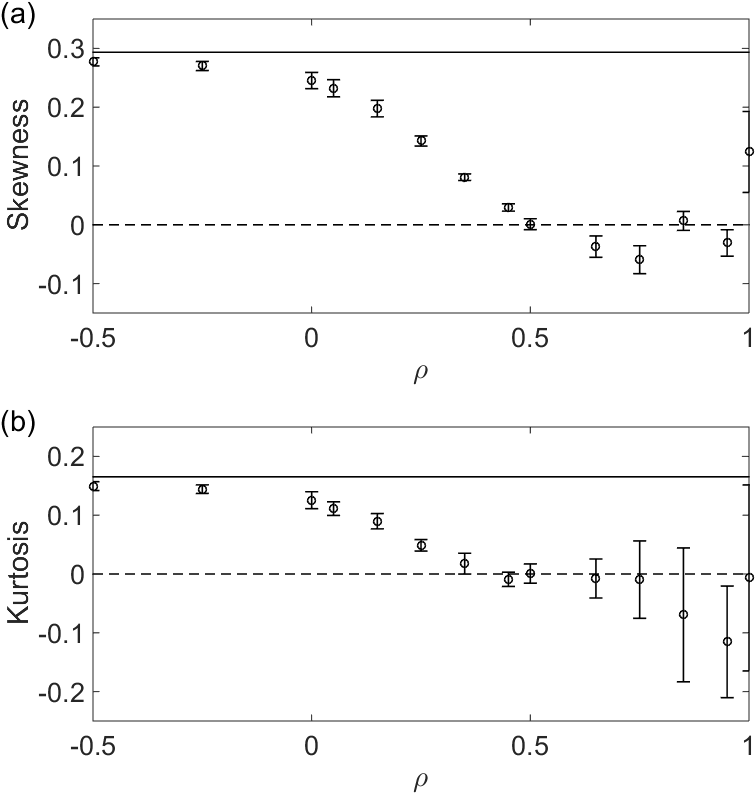}
\caption{{(a)} Skewness and {(b)} kurtosis for DPRM with spatially correlated noise, compared to the TW-GOE values 
(solid lines), $s = 0.293$ and $k = 0.165$, respectively: Both approach the TW-GOE values for 
$\rho \leq 0$, and decrease to 0 as $\rho$ increases to 1/2. 
Beyond $\rho = 1/2$, the uncertainties are too large to rule out $s=k=0$.}
\label{fig: sk}
\end{figure}

The skewness $s$ and kurtosis $k$,  plotted in Fig. \ref{fig: sk}, are obtained by averaging results 
over the above snapshots in $t$.
In the uncorrelated case, it is possible to estimate the true asymptotic values of $s$ and $k$ 
using methods developed in Ref.~\cite{km90}. 
In the correlated case, however, we run into problems as the uncertainties grow rapidly with correlation.
For $\rho \leq 0$, the skewness and kurtosis approach those of the TW-GOE, 
the limiting distribution for uncorrelated noise. 
As $\rho$ increases towards 1/2, both $s$ and $k$ decrease to 0, and the distribution becomes more symmetric. 
In particular, the data suggests that $s, k \rightarrow 0$ as $\rho \rightarrow 1/2$, 
consistent with a Gaussian distribution. 
This would be expected if $\rho=1/2$ corresponded to random energies fully correlated in the $x$-direction, 
but randomly changing along the $t$-direction. 
The energy of the DPRM would then be a sum of random variables, thus $\beta=1/2$, 
while the path executes a random walk with $z=2$.
The latter is not correct, as the Galilean exponent identity at $\beta = 1/2$ leads to the numerically observed exponent of 
$z = 4/3$. 
We note also that the Fourier transform procedure for generating spatially correlated noise, 
devised in Ref.~\cite{mhss96} and used here, actually produces correlations which vary logarithmically 
at $\rho = 1/2$,  as $\langle \eta(x,t) \eta(x',t') \rangle \sim (a-b \ln |x - x'|) \delta(t - t')$ (see Appendix).
Interestingly the RG result (also based on Fourier transformed noise) also predicts the observed $\beta=1/2$.
Finally, for $\rho > 1/2$, the uncertainty in $s$ and $k$ grow rapidly due to the increased correlations in random energies, 
and we cannot conclusively state whether the distribution is Gaussian or not.
It is of interest to note that another class of distributions, interpolating between TW-GUE and Gaussian, was found in Ref.~\cite{j07} in the context of random matrix theory, and the convergence of TW distributions to the Gumbel distribution was studied in Ref.~\cite{ad14}.

In summary, we have performed extensive simulations of the DPRM over a spatially correlated landscape. 
The energy fluctuations are well described by scaling exponents $\beta$ and $z$, 
extracted by standard data collapse.
We find that the exponent identity  $2/z - \beta = 1$ reflecting Galilean invariance holds for all $\rho$,
while the related identity $\chi + z = 2$ is apparently violated for $\rho \geq 1/2$, most likely due to
systematic errors.
While the growth exponent $\beta(\rho)$ is qualitatively similar to predictions of the RG, 
there are significant deviations for $\rho<1/2$.
In particular, for the important value of $\rho=1/2$, at the borderline of correlations
growing or decaying with separation, there is strong evidence that $\beta=1/2$, with Gaussian
energy fluctuations.

For an uncorrelated landscape, the optimal energy of DPRM behaves as 
$E = f_\infty t + (\Gamma t)^{1/3} \xi$, where $f_\infty$ and $\Gamma$ are non-universal, system-dependent parameters, and $\xi$ is a $\mathcal O(1)$ random variable obeying TW-GOE statistics. 
There is currently no analytical prediction for the limiting distribution in the case of correlated noise.
From the overall scaling, we can propose an analogous form, 
$E = f_\infty t + (\Gamma t)^{\beta(\rho)} \xi(\rho)$, where $\beta(\rho)$ is the modified growth exponent. 
The random variable $\xi(\rho)$ is distributed according to TW-GOE statistics for $\rho \leq 0$.
A priori one could have imagined that the distribution retains the TW form in general, or that it discontinuously
transitions to a different distribution for $\rho>0$.
Instead, we observe a smooth shift as $\rho$ increases, to a Gaussian form at $\rho = 1/2$. 
For $\rho > 1/2$, the uncertainty in skewness and kurtosis become too significant to conclude 
whether the distribution is Gaussian. 
We thus find a  class of distributions, interpolating between TW and Gaussian, 
which governs the statistics of DPRM with spatially correlated noise.

We are grateful to T.~Halpin-Healy, A.~Rosso, and G.~Bunin for illuminating discussions,
and financial support from the NSF through grant No. DMR-12-06323. This work was
partly completed at KITP, which is supported
in part by the NSF under grant PHY05-51164. 
SC was also supported by MIT through the Thomas Frank Fellowship.

\section*{Appendix: Algorithm for generating correlated noise}

\renewcommand\theequation{A\arabic{equation}}
\setcounter{equation}{0}

We present here the method used for generating correlated noise developed in Ref.~\cite{mhss96}.
Note that a similar method was developed in Ref.~\cite{pyhh95} to study DPRM and BD models with correlated noise.

\renewcommand\thefigure{A\arabic{figure}}
\setcounter{figure}{0}

\begin{figure}
\includegraphics[scale=0.385]{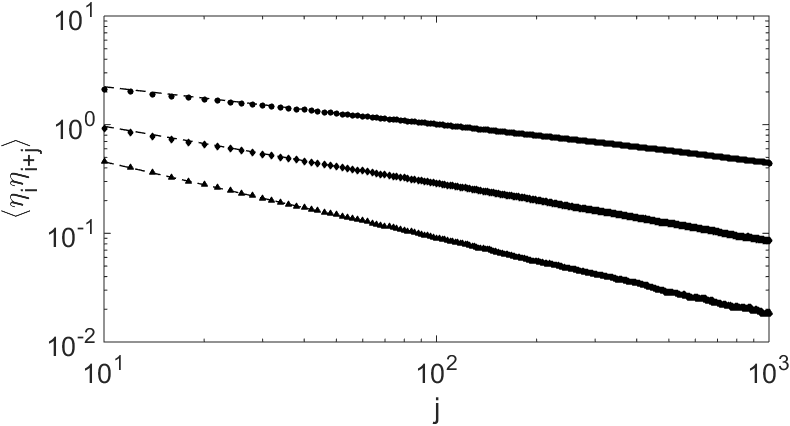}
\caption{Log-log plot of correlation of generated noise with $\rho < 1/2$ for system size $L=10^6$. The data for $\rho = 0.35$, $\rho = 0.25$, and $\rho = 0.15$ are plotted (from top to bottom). The best fit lines (dashed) give $\rho = 0.33\pm0.02$, $\rho = 0.24\pm0.02$, and $\rho = 0.15\pm0.01$ respectively.}
\label{fig:corr1}
\end{figure}

\begin{figure}
\includegraphics[scale=0.385]{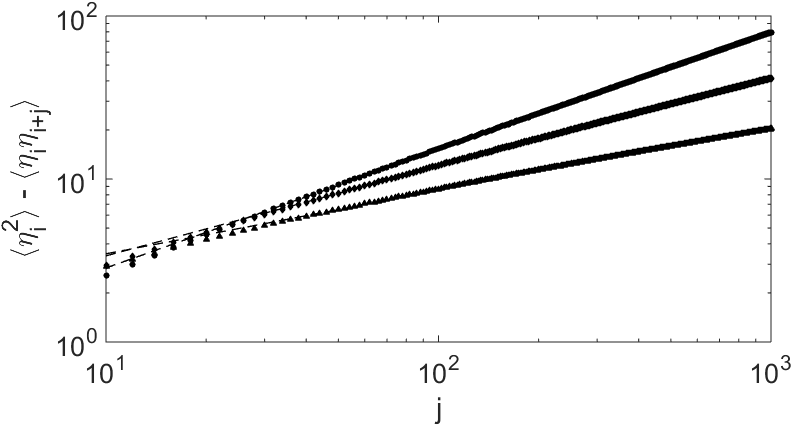}
\caption{Log-log plot of correlation of generated noise with $\rho > 1/2$ for system size $L=10^6$. The data for $\rho = 0.85$, $\rho = 0.75$, and $\rho = 0.65$ are plotted (from top to bottom). The best fit lines (dashed) give $\rho = 0.86\pm0.03$, $\rho = 0.77\pm0.04$, and $\rho = 0.69\pm0.04$ respectively.}
\label{fig:corr2}
\end{figure}

\begin{figure}
\includegraphics[scale=0.385]{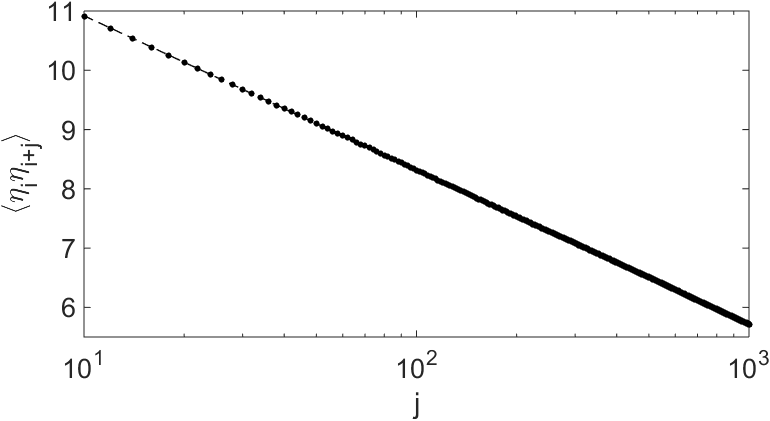}
\caption{Semilog plot of correlation of generated noise with $\rho = 1/2$ for system size $L=10^6$. The correlation decays logarithmically with separation, and the best fit line (dashed) gives $C(j) = -1.13\ln|j|+13.52$.}
\label{fig:corr3}
\end{figure}

In $d=1$ dimension, the goal is to use a sequence of Gaussian i.i.d. random variables $\{ u_i \}_{i=1, \ldots, L}$ to generate a sequence $\{ \eta_i \}_{i=1,\ldots, L}$ with correlation function~\cite{mhkz89}
\begin{equation}\label{eq:corr func def}
C(j) = \langle \eta_i \eta_{i+j} \rangle \sim j^{2\rho - 1}, \tab j \rightarrow \infty.
\end{equation}
Taking the Fourier transform gives the spectral density function $S(q)$, which has asmptotic form
\begin{equation}\label{eq:spec den def}
S(q) = \langle \eta_q \eta_{-q} \rangle \sim q^{-2\rho}, \tab q \rightarrow 0.
\end{equation}
The correlated random variables can then be obtained from the relation
\begin{equation}\label{eq:etaq}
\eta_q = [ S(q) ]^{1/2} u_q,
\end{equation}
where $\{ \eta_q \}$ and $\{ u_q \}$ are the Fourier transform coefficients of $\{ \eta_i \}$ and $\{ u_i \}$, respectively.

As in Ref.~\cite{mhss96}, we define the correlation function to be
\begin{equation}\label{eq:mod corr func}
C(j) \equiv (1+j^2)^{\rho - 1/2},
\end{equation}
on the interval $j \in [-L/2, L/2]$. This then has the same asymptotic power law decay as in Eq.~(\ref{eq:corr func def}). The spectral density can be calculated analytically as the discrete Fourier transform,
\begin{equation}\label{eq:mod spec den}
S(q) = \frac{2\pi^{1/2}}{\Gamma(- \rho+1)} \left( \frac{q}{2} \right)^{-\rho} K_{\rho}(q),
\end{equation}
where $q = 2\pi n/L$ with $n = - L/2, \ldots, L/2$, and $K_\rho$ is the modified Bessel function of the second kind of order $\rho$. We define $S(q = 0) = 0$ to avoid any divergences. For $\rho > 0$, the modified Bessel function has asymptotic form
\begin{equation}\label{eq:bessel}
K_\rho(q) \sim \left\{ \def\arraystretch{1.5} \begin{array}{ll} \displaystyle \frac{\Gamma(\rho)}{2} \left( \frac{2}{q} \right)^\rho, & q \ll 1, \\
\displaystyle \sqrt \frac{\pi}{2q} e^{-q}, & q \gg 1. \end{array} \right.
\end{equation}
As $q \rightarrow 0$, we recover the asymtotic form in Eq.~(\ref{eq:spec den def}).

The process for generating the correlated noise used to study DPRM can then be summarized in the following steps.
\begin{enumerate}
\item Generate Gaussian i.i.d. random variables $\{ u_i \}$, and calculate $\{ u_q \}$ using the fast Fourier transform.
\item Calculate the spectral density function $S(q)$ using Eq.~(\ref{eq:mod spec den}),~(\ref{eq:bessel}).
\item Calculate $\{ \eta_q \}$ using Eq.~(\ref{eq:etaq}), and calculate $\{ \eta_i \}$ using the inverse Fourier transform.
\end{enumerate}

For the DPRM model we simulated, the system size is $L=10^6$, evolved over $t=10^4$ time steps. We check the correlation of the noise generated using the above method by averaging over $10^2$ realizations. The results for $\rho < 1/2$ and $\rho > 1/2$ are plotted in Fig.~\ref{fig:corr1} and~\ref{fig:corr2}, respectively, and we see that the data is in good agreement with the expected values of $\rho$ up to a separation of $j = 10^3$. In the special case $\rho = 1/2$, we find that the correlation decays logarithmically with separation, as shown in Fig.~\ref{fig:corr3}.


\begin{thebibliography}{99}

\bibitem{tw96}
C. A. Tracy and H. Widom, Commun. Math. Phys. {\bf 177}, 727 (1996).

\bibitem{quanta}
N. Wolchover, Quanta Magazine, 10/27 (2014).

\bibitem{mn05}
S. N. Majumdar and S. Nechaev, Phys. Rev. E {\bf 72}, 020901 (R) (2005).

\bibitem{Ch12}
Ch. Hajiyev, ISA Transactions, {\bf 51}, 189 (2012).

\bibitem{kz87}
M. Kardar and Y.-C. Zhang, Phys. Rev. Lett. {\bf 58}, 2087 (1987).

\bibitem{hhz94}
T. Halpin-Healy and Y.-C. Zhang, Phys. Rep. {\bf 254}, 215 (1995).

\bibitem{hht15}
T. Halpin-Healy and K. A. Takeuchi, J. Stat. Phys. {\bf 160}, 794 (2015).

\bibitem{ic12}
I. Corwin, Random Matrices: Theory Appl. 1, 1130001 (2012).

\bibitem{qs15}
J. Quastel and H. Spohn, J. Stat. Phys. {\bf 160}, 965 (2015).

\bibitem{mhkz89}
E. Medina, T. Hwa, M. Kardar, and Y.-C. Zhang,
Phys. Rev. A {\bf 39}, 3053 (1989).

\bibitem{kpz86}
M. Kardar, G. Parisi, and Y.-C. Zhang, Phys. Rev. Lett. {\bf 56}, 889 (1986).

\bibitem{kmb91a}
J. M. Kim, M. A. Moore, and A. J. Bray, Phys. Rev. A {\bf 44}, 2345 (1991).

\bibitem{kmb91b}
J. M. Kim, M. A. Moore, and A. J. Bray, Phys. Rev. A {\bf 44}, R4782 (1991).

\bibitem{hh91}
T. Halpin-Healy, Phys. Rev. A {\bf 44}, R3415 (1991).

\bibitem{hh12-13}
T. Halpin-Healy, Phys. Rev. Lett. {\bf 109} 170602 (2012); Phys. Rev. E {\bf 88}, 042118 (2013).

\bibitem{hhl14}
T. Halpin-Healy and Y. Lin, Phys. Rev. E {\bf 89}, 010103(R) (2014).

\bibitem{gldbr}
T. Gueudre, P. Le Doussal, J.-P. Bouchaud, and A. Rosso, Phys. Rev. E {\bf 91} 062110 (2015).

\bibitem{e61}
M. Eden, {\it Proc. 4th Berkeley Symp. on Mathematical Statistics and Probability} (University of California Press 1961).

\bibitem{m87}
P. Meakin, J. Phys. A: Math. Gen. {\bf 20}, L1113 (1987).

\bibitem{rhh91}
S. Roux, A. Hansen, and E. L. Hinrichsen, J. Phys. A: Math. Gen. {\bf 24} L295 (1991).

\bibitem{kk89}
J. M. Kim and J. M. Kosterlitz, Phys. Rev. Lett. {\bf 62}, 2289 (1989).

\bibitem{kkal91}
J. M. Kim, J. M. Kosterlitz, and T. Ala-Nissila, J. Phys. A: Math. Gen. {\bf 24} 5569 (1991).

\bibitem{v59}
M. J. Vold, J. Colloid Sci. {\bf 14}, 168 (1959).

\bibitem{k87}
M. Kardar, Nucl. Phys. B{\bf290} [FS20], 582 (1987).

\bibitem{ss10}
T. Sasamoto, and H. Spohn, Phys. Rev. Lett. {\bf 104}, 230602 (2010).

\bibitem{j00}
K. Johansson, Commun. Math. Phys. {\bf 209}, 437 (2000).

\bibitem{ps00}
M. Pr\"ahofer and H. Spohn, Phys. Rev. Lett. {\bf 84}, 4882 (2000).

\bibitem{z90}
Y.-C. Zhang, Physics A: Statistical Mechanics and its Applications, {\bf 170}, 1 (1990).

\bibitem{jk91}
J. Krug, J. de Phys. I, {\bf 1}, 9 (1991).

\bibitem{bbp07}
G. Biroli, J.-P. Bouchaud, and M. Potters, J. Stat. Mech.: Theory and Experiment, {\bf 2007}, P07019 (2007).

\bibitem{fns77}
D. Forster, D. R. Nelson, and M. J. Stephen, Phys. Rev. A {\bf 16}, 732 (1977).

\bibitem{ftj99}
E. Frey, U. C. T\"auber, and H. K. Janssen, Europhys. Lett. {\bf 47},  14 (1999).

\bibitem{kcdw14}
T. Kloss, L. Canet, B. Delamotte, and N. Wschebor, Phys. Rev. E {\bf 89}, 022108 (2014).

\bibitem{mhss96}
H. A. Makse, S. Havlin, M. Schwartz, H. E. Stanley, Phys. Rev. E {\bf 53}, 5445 (1996).

\bibitem{pyhh95}
N.-N. Pang, Y.-K. Yu, and T. Halpin-Healy, Phys. Rev. E {\bf 52}, 3224 (1995).

\bibitem{alf91}
J. G. Amar, P.-M. Lam, and F. Family, Phys. Rev. A {\bf 43}, 4548 (1991).

\bibitem{phss91}
C.-K. Peng, S. Havlin, M. Schwartz, and H. E. Stanley, Phys. Rev. A {\bf 44}, R2239 (1991).

\bibitem{wb95}
M. Wu and K. Y. R. Billah, Phys. Rev. E {\bf 51}, 995 (1995).

\bibitem{km90}
J. Krug and P. Meakin, J. Phys. A: Math. Gen. {\bf 23}, L987 (1990).

\bibitem{j07}
K. Johansson, Probab. Theory Related Fields {\bf 138}, 75 (2007).

\bibitem{ad14}
R. Allez and L. Dumaz, J. Stat. Phys. {\bf 156}, 1146 (2014).



%
%
%

\end{thebibliography}


\end{document}